\documentclass[final,5p,times, sort&compress, fleqn,twocolumn]{elsarticle}
\journal{Journal of Magnetism and Magnetic Materials}

\usepackage{amsmath}

\begin{document}

\begin{frontmatter}
\title{Non-chiral spin frustration versus highly degenerate ferromagnetic state with local chiral degrees of freedom of an exactly solvable spin-electron planar model \\ of~inter-connected trigonal bipyramids\tnoteref{mytitlenote}}
   
\tnotetext[mytitlenote]{The work was financially supported by the Slovak Research and Development Agency under the contract No. APVV-16-0186 and by the Ministry of Education, Science, Research and Sport of the Slovak Republic under the grants Nos. KEGA 021TUKE-4/2020 and VEGA 1/0301/20.}

\author{Lucia G\'alisov\'a}
\address{Institute of Manufacturing Management, 
  	     Faculty of Manufacturing Technologies with the seat in Pre\v{s}ov, Technical University of~Ko\v{s}ice, \\
  	     Bayerova 1, 080\,01 Pre\v{s}ov, Slovakia}

\ead{lucia.galisova@tuke.sk}

\begin{abstract}
The frustration phenomenon in an exactly solvable spin-electron planar model constituted by identical bipyramidal plaquettes is discussed within the Toulouse's and dos Santos and Lyra's frustration concepts. It is shown that the ground state of the model contains the unfrustrated spontaneously ordered quantum ferromagnetic phase with local chiral degrees of freedom in the electron sub-lattice and the disordered quantum one, where both the Ising and electron sub-lattices are frustrated. The frustration of the latter sub-lattice persists at finite temperatures, but only in the disordered region. It finally vanishes at a certain frustration temperature. The reentrant behaviour of the frustration in the electron sub-lattice with three consecutive frustration temperatures due to a competition with the unfrustrated ferromagnetic spin arrangement near the ground-state phase transition can also be observed.
\end{abstract}


\end{frontmatter}

\section{Introduction}
\label{sec:1}

The examination of frustrated magnetic systems is dated back to 1950s, when it was found that the Ising antiferromagnetic triangular lattice has quite different properties to ferromagnets or bipartite antiferromagnets~\cite{Wan50,Hou50}. Since that time the frustration phenomenon has been a topic of constant scientific interest. Thanks to the relentless desire of theoretists and experimentalists to clarify the unusual and discover the unknown, the area of frustrated magnetism has been expanded considerably over the last two decades. Many current studies on frustrated magnetic systems are simultaneously focused on other phenomena with more degrees of freedom, such as magneto-elastic couplings, dilution effects, orbital degrees of freedom, or electron doping (see, e.g., Ref.~\cite{Lac11} and references therein).  

Recent studies show that the exactly solvable pure Ising and also mixed Ising-Heisenberg models on decorated planar lattices involving triangular structures represent suitable playground for a comprehensive rigorous investigation of the frustration phenomenon in two dimensions (2D)~\cite{Str13,Str15,Gal19}. However, these studies bring no insight into the frustration of mobile particles.  
To fill this gap, in this paper a mixed spin-electron model on a regular 2D lattice consisting of identical inter-connected trigonal bipyramids will be explored in connection to the expected frustration therein. Apart the original Toulouse's concept of frustration, which calls a spin frustrated when it cannot take a state to satisfy all the exchange interactions with its spin neighbours~\cite{Tol77},  
the alternative definition introduced by R.~J.~V. dos Santos and M.~L. Lyra, determining the frustrated state by a negative value of the product of pair correlation functions along the elementary plaquette~\cite{San92}, will be used to investigate the phenomenon.  
The latter concept is temperature dependent, which allows one to examine the phenomenon even near continuous phase transitions of the system. 

The outline of the paper is as follows. In Sec.~\ref{sec:2} the model is introduced and crucial steps of its exact solution resulting in the exact closed-form expression for the partition function are briefly mentioned. Sec.~\ref{sec:3} contains the most interesting numerical results. Finally, the basic findings are summarized in Sec.~\ref{sec:4}.

\section{Model and its exact treatment}
\label{sec:2}

\begin{figure}[t!]
	\centering
	\includegraphics[width=0.95\columnwidth]{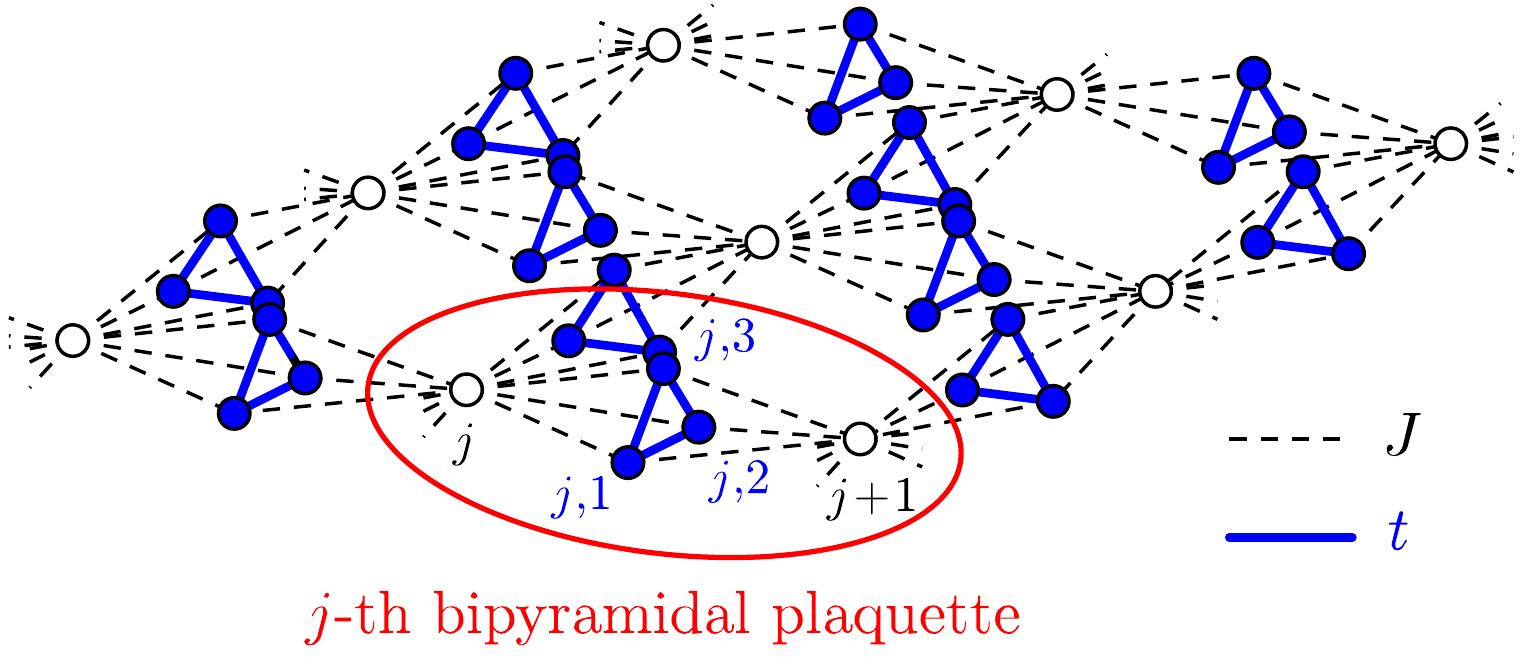}
	\vspace{0mm}
	\caption{The regular 2D lattice of the mixed spin-electron trigonal bipyramidal plaquettes. White circles illustrate nodal lattice sites occupied by the Ising spins $\mu=1/2$, while the blue ones are available for mobile electrons. Dashed black lines label the Ising-type exchange interaction $J$ and the solid blue ones label the skipping paths $t$ of mobile electrons.}
	\label{fig:1}
\end{figure}

Let us consider the mixed spin-electron model on a regular 2D lattice which is schematically depicted in Fig.~\ref{fig:1}. The common vertices of the plaquettes (white circles), which represent nodal lattice sites, are occupied by the localized Ising spins of the magnitude $1/2$, while the rest ones (blue circles), forming triangles oriented perpendicularly to the plaquette's axes, are available for mobile electrons. Our attention will be focused on the model with two electrons are delocalized over the triangular cluster of each bipyramidal plaquette. The electron hopping between different plaquettes is forbidden. 
Assuming $N$ nodal lattice sites ($N\to\infty$), the total Hamiltonian of the model can be written as a sum of $2N$ commuting plaquette Hamiltonians: 
\begin{equation}
 \hat{H} = \sum_{j=1}^{2N}\hat{H}_j,   
\end{equation}
where $\hat{H}_j$ involves all interactions realized within the corresponding bipyramidal plaquette:
\begin{eqnarray}
\label{eq:H_j}
\hat{H}_j\!\!\!\!&=&\!\!\!\! -t\!\!\sum_{\sigma \in\{\uparrow, \downarrow\}}\sum_{k=1}^3 \big(\hat{c}_{j,k,\sigma}^{\dag}\hat{c}_{j,k+1,\sigma} \!+ {\rm h.c.}\big)
\nonumber\\
\!\!\!\!&&\!\!\!\! -J\hat{S}_{\!j}^z\big(\hat{\mu}_{j}^{z} + \hat{\mu}_{j+1}^{z}\big) + U \!\sum_{k = 1}^{3}\hat{n}_{j,k,\uparrow}\hat{n}_{j,k,\downarrow}.
\end{eqnarray}
In above, $\hat{c}_{j,k,\sigma}^{\dag}$ ($\hat{c}_{j,k,\sigma}$) represents fermionic creation (annihilation) operator for mobile electrons with the spin $\sigma \in\{ \uparrow, \downarrow\}$ at the $k$-th site of the $j$-th triangular cluster, $\hat{n}_{j,k,\sigma} = \hat{c}_{j,k,\sigma}^{\dag}\hat{c}_{j,k,\sigma}$ is the fermion number operator,
$\hat{S}_{\!j}^{z} = \sum_{k=1}^3\hat{S}_{\!j,k}^z = \sum_{k=1}^3(\hat{n}_{j,k,\uparrow} - \hat{n}_{j,k,\downarrow})/2$ labels the total spin of the $j$-th electron triangle in $z$-direction, and $\hat{\mu}_{j}^{z}$ is the $z$-component of the spin-$1/2$ operator for the Ising spin at the $j$-th nodal site. The term $t>0$ represents the hopping parameter taking into account the kinetic energy of mobile electrons, $J$ labels the Ising-type interaction between the Ising spins and their nearest electron neighbours, and $U>0$ represents the Coulomb repulsion between electrons at the same lattice site. For the sake of simplicity of calculations, the periodic boundary conditions $\hat{c}_{j, 4,\sigma}^{\dag} \equiv \hat{c}_{j,1,\sigma}^{\dag}$ ($\hat{c}_{j,4,\sigma} \equiv \hat{c}_{j,1,\sigma}$) and $\hat{\mu}_{2N+1}^{z} \equiv \hat{\mu}_{1}^{z}$ are assumed for the operators. %

The plaquette Hamiltonian~(\ref{eq:H_j}) has in total fifteen eigenvalues, some of which are two-fold degenerate:
\begin{subequations}
\begin{eqnarray}
\label{eq:E_1-5}
E_{1,2} \!\!\!\!&=&\!\!\!\!  -t -J(\mu_{j}^{z} + \mu_{j+1}^{z}),
\quad
E_{3,4} = -t, \quad
E_{5} = 2t, 
\\
\label{eq:E_6-9}
E_{6,7} \!\!\!\!&=&\!\!\!\! -t + J(\mu_{j}^{z} + \mu_{j+1}^{z}),
\quad
E_{8,9} = 2t
\pm J(\mu_{j}^{z} + \mu_{j+1}^{z}), \hspace{1.0cm}
\\
\label{eq:E_10,11}
E_{10,11} \!\!\!\!&=&\!\!\!\! \frac{U}{2} + \frac{t}{2} + \frac{1}{2}\!\sqrt{U^2\!-\!2Ut\!+\!9t^2}\!,
\\
\label{eq:E_12,13}
E_{12,13} \!\!\!\!&=&\!\!\!\! \frac{U}{2} + \frac{t}{2} - \frac{1}{2}\!\sqrt{U^2\!-\!2Ut\!+\!9t^2}\!,
\\
\label{eq:E_14,15}
E_{14,15} \!\!\!\!&=&\!\!\!\! \frac{U}{2} - t \pm \frac{1}{2}\!\sqrt{U^2\!+\!4Ut\!+\!36t^2}\!.
\end{eqnarray}
\end{subequations}
The set of eigenvalues~(\ref{eq:E_1-5})--(\ref{eq:E_14,15}) can be employed for a ground-state analysis as well as a calculation of the partition function $Z$ of the model, which is partially factorizable and writable in terms of~(\ref{eq:E_1-5})--(\ref{eq:E_14,15}) due to commutative character of $\hat{H}_j$:
\begin{equation}
\label{eq:Z}
Z = \mathrm{Tr}\,{\rm e}^{-\beta\hat{H}} = \sum_{\{\mu_j^z\}}\prod_{j=1}^{2N}\mathrm{Tr}_{j}\,{\rm e}^{-\beta\hat{H}_j} = \sum_{\{\mu_j^z\}}\prod_{j=1}^{2N}\sum_{n=1}^{15} \,{\rm e}^{-\beta E_{n}}.
\end{equation}
In above, $\beta = 1/(k_{\rm B}T)$ ($k_\mathrm{B}$ is the Boltzmann's constant, $T$ is the absolute temperature of the system), $\sum_{\{\mu_j^z\}}$ runs over all possible spin states of the nodal Ising spins, $\prod_{j=1}^{2N}$ runs over all bipyramidal plaquettes (one-third filled electron triangular clusters), and $\mathrm{Tr}_{j}$ stands for the trace over all possible degrees of freedom of the electron pair in the $j$-th triangular cluster.

Thanks to its specific lattice topology, the spin-electron model under consideration can be included into a wide class of bond-decorated Ising planar lattices, which are exactly solvable by means of the algebraic technique named generalized decoration-iteration mapping transformation~\cite{Fis59,Str10}. As a result, one obtains the rigorous correspondence between the partition function $Z$ of the considered model and the known partition function $Z_{IM}$ of the uniform spin-$1/2$ Ising square lattice~\cite{Ons44}: 
\begin{equation}
\label{eq:ZZI}
Z(\beta, t, J, U) = A^{2N}Z_{IM}(J_{eff}).
\end{equation}
The parameters $A$ and $J_{eff}$ present in Eq.~(\ref{eq:ZZI}) are the temperature-dependent functions of the model's parameters $t$, $J$, $U$. Their explicit expressions can be obtained from the ’self-consistency’ condition of the used algebraic approach. For more computational details we refer the reader to, e.g., Ref.~\cite{Str09}, which deals with similar 2D model. We note that the mapping relation~(\ref{eq:ZZI}) represents the central result of our calculations, because it is an indispensable part of the rigorous calculation  of all important physical quantities.

\section{Results and discussion}
\label{sec:3}

In this section, we proceed to a discussion of the most interesting numerical results for the considered spin-electron model. As the sign change $J\leftrightarrow-J$ of the Ising-type interaction between the Ising spins and mobile electrons does not cause any fundamental difference in the magnetic behaviour of the system, we can restrict our attention to its particular version with the ferromagnetic coupling $J>0$ without loss of generality.

\subsection{Spin frustration in the ground state}
\label{subsec:1}

\begin{figure}[t!]
	\centering
	\vspace{0mm}
	\includegraphics[width=0.95\columnwidth]{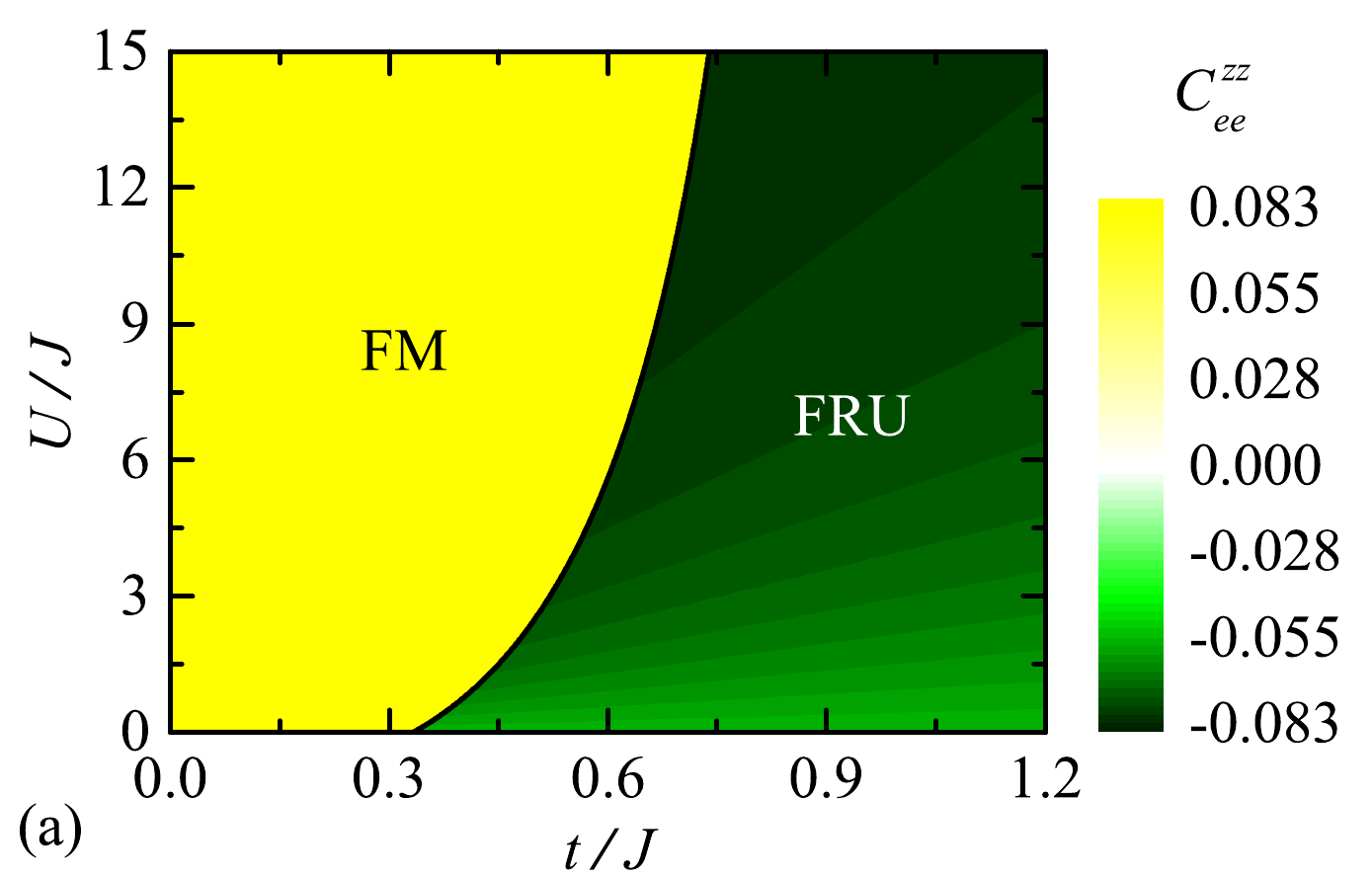}
	\includegraphics[width=0.95\columnwidth]{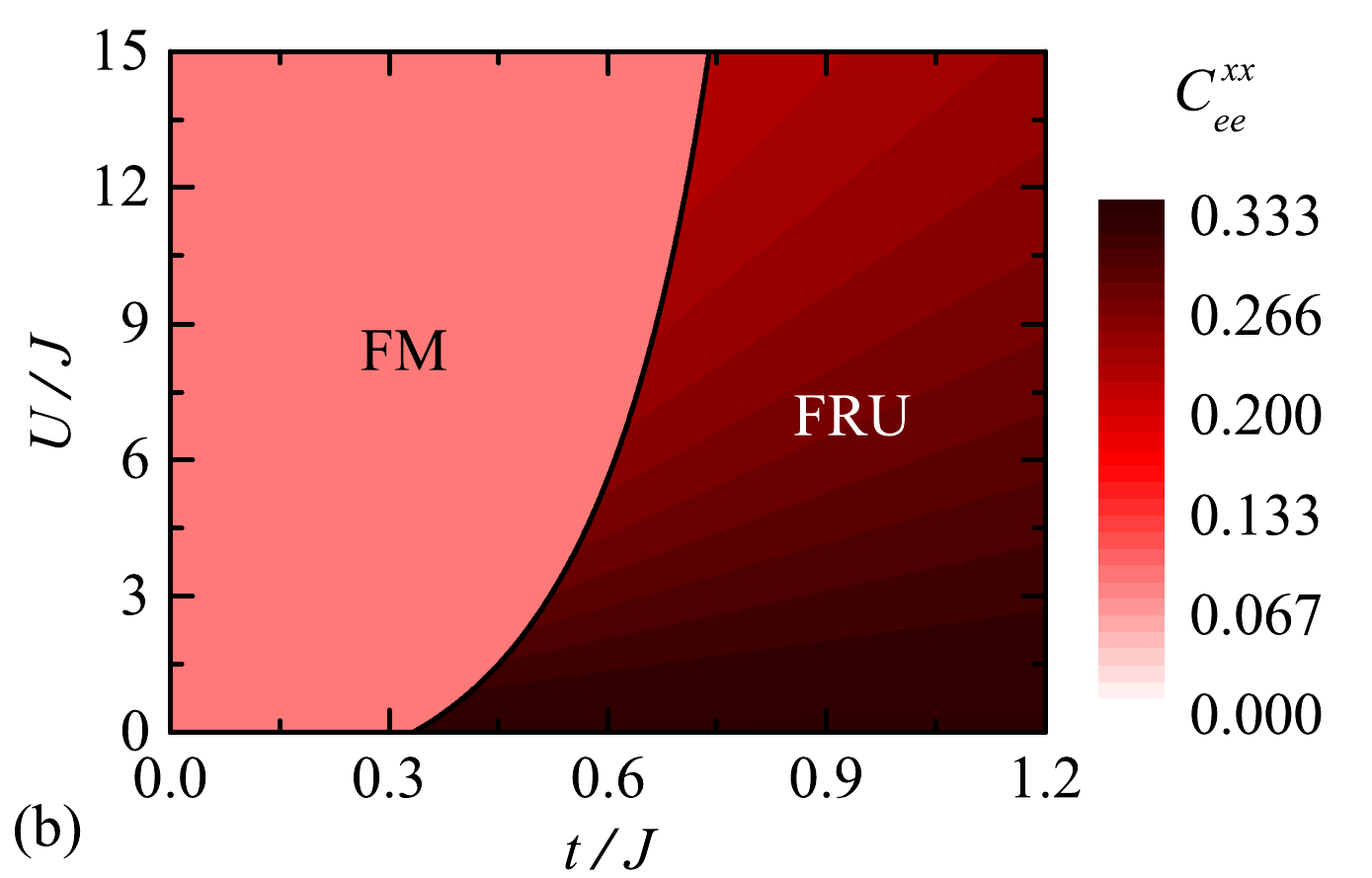}
	\vspace{-2mm}
	\caption{The ground-state phase diagram of the studied spin-electron model in the $t/J-U/J$ parameter plane supplemented by density plots of the pair correlation functions specifying the spin ordering between two electrons in a given triangular cluster along $z$-axis [panel (a)] and along $x$-axis [panel (b)].}
	\label{fig:2}
\end{figure}

First we take a closer look at magnetic ground-state arrangement of the model. A systematic comparison of the eigenvalues~(\ref{eq:E_1-5})--(\ref{eq:E_14,15}) of the plaquette Hamiltonian~(\ref{eq:H_j}) for all spin-state combinations of the Ising pair $\mu_{j}^{z}$, $\mu_{j+1}^{z}$ reveals that the zero-temperature phase diagram of the studied model includes two macroscopically degenerate quantum phases. One is the spontaneously long-range ordered ferromagnetic (FM) phase with the energy $E_{\rm FM} = -2N(J + t)$, which is characterized by the total spin $S_{\!j}^z=1$ of the electron triangles due to a ferromagnetic spin arrangement of the electron pairs in them and a parallel alignment of the nodal Ising spins to their electron neighbours:
\begin{equation}
\label{eq:FM}
|{\rm FM}\rangle = \prod_{j=1}^{2N}
        |\uparrow\rangle_{\mu_j^z}\otimes \big|S_{\!j}^z=1, R \textrm{ or } L\big\rangle_{\!\triangle_j}.
\end{equation}
The eigenvector $|S_{\!j}^z = 1, R \textrm{ or } L\rangle_{\triangle_j}$ in Eq.~(\ref{eq:FM}) contains two opposite chiral degrees of freedom of the $j$-th electron triangular cluster, namely $R$ight- and $L$eft-hand side one:
\begin{eqnarray*}
\big|S_{\!j}^z=1, R\big\rangle_{\!\triangle_j} \!\!\!\!\!&=&\!\!\!\!  \frac{1}{\sqrt{3}}\left(|\uparrow,\uparrow,\emptyset\rangle_{\triangle_j}\!+{\rm e^{\frac{2\pi{\rm i}}{3}}}|\emptyset,\uparrow,\uparrow\rangle_{\triangle_j} \!+{\rm e^{\frac{4\pi{\rm i}}{3}}}|\uparrow,\emptyset,\uparrow\rangle_{\triangle_j}\right), \\
\big|S_{\!j}^z=1, L\big\rangle_{\!\triangle_j} \!\!\!\!\!&=&\!\!\!\!  \frac{1}{\sqrt{3}}\left(|\uparrow,\uparrow,\emptyset\rangle_{\triangle_j}\!+{\rm e^{\frac{4\pi{\rm i}}{3}}}|\emptyset,\uparrow,\uparrow\rangle_{\triangle_j} \!+{\rm e^{\frac{2\pi{\rm i}}{3}}}|\uparrow,\emptyset,\uparrow\rangle_{\triangle_j}\right), \end{eqnarray*}
which is reflected in the residual entropy ${\cal S}/(2Nk_{\rm B}) = \ln 2\approx0.693$ per bipyramidal unit cell.
Other phase is the disordered frustrated (FRU) phase 
with the energy $E_{\rm FRU} = N\big(U-2t-\!\sqrt{U^2+4Ut+36t^2}\big)$, where the nodal Ising spins are free to flip between up and down states and the total spin $S_{\!j}^z$ of the electron triangles is zero:
\begin{equation}
\label{eq:FRU}
|{\rm FRU}\rangle = \prod_{j=1}^{2N}
        |\uparrow(\downarrow)\rangle_{\mu_j^z}\otimes \big|S_{\!j}^z=0\big\rangle_{\!\triangle_j}.
\end{equation}
The total spin $S_{\!j}^z = 0$ is a result of a quantum superposition of six intrinsic antiferromagnetic and three non-magnetic ionic states of electron pairs:
\begin{eqnarray*}
\big|S_{\!j}^z=0\big\rangle_{\!\triangle_j} \!\!\!\!\!&=&\!\!\!\!  \frac{\sin\varphi}{\sqrt{6}}\left(|\uparrow,\downarrow,\emptyset\rangle_{\triangle_j}\!+|\emptyset,\uparrow,\downarrow\rangle_{\triangle_j} \!+|\downarrow,\emptyset,\uparrow\rangle_{\triangle_j}\right. \\
&&\!\!\!\!
\left.
-\,|\downarrow,\uparrow,\emptyset\rangle_{\triangle_j}\!-|\emptyset,\downarrow,\uparrow\rangle_{\triangle_j} \!-|\uparrow,\emptyset,\downarrow\rangle_{\triangle_j}\right)
\\
&&\!\!\!\!
+\frac{\cos\varphi}{\sqrt{3}}\left(|\uparrow\downarrow,\emptyset,\emptyset\rangle_{\triangle_j}\!+|\emptyset,\uparrow\downarrow,\emptyset\rangle_{\triangle_j} \!+|\emptyset,\emptyset,\uparrow\downarrow\rangle_{\triangle_j}\right), 
\end{eqnarray*}
where $\tan\varphi = \sqrt{2}\big( U+2t+\!\sqrt{U^2+4Ut+36t^2}\big)/(8t)$. Interestingly, the spin frustration of the Ising sub-lattice leads to a half lower value of the residual entropy ${\cal S}/(2Nk_{\rm B}) = \ln 2^{1/2}\approx0.347$ of the FRU phase compared to that which can be identified in the previous FM ground state.
The stability of the spontaneously ordered FM phase is limited to the relatively small values of the hopping term $t/J< \big[\!\sqrt{(U/J + 6)^2 + 24U/J} - U/J\big]/18$. Otherwise, the disordered FRU phase can be identified as the ground state (see Fig.~\ref{fig:2}). 

The character of the spin-electron arrangement in the individual phases is consistent with the corresponding zero-temperature asymptotic values of the spontaneous sub-lattice magnetization $m_I = \big\langle\hat{\mu}_j^z\big\rangle$ per nodal Ising spin and $m_{e} = \big\langle \hat{S}_{\!j}^z\big\rangle$ per electron triangle of a given plaquette, as well as the corresponding values and/or analytical expressions of the pair correlation functions $C_{Ie}^{zz}=\big\langle \hat{\mu}_{j}^z\hat{S}_{\!j,k}^z\big\rangle$, $C_{ee}^{zz}=\big\langle \hat{S}_{\!j,k}^z\hat{S}_{\!j,k+1}^z\big\rangle$, $C_{ee}^{xx}=\big\langle \hat{S}_{\!j,k}^x\hat{S}_{\!j,k+1}^x\big\rangle$, which are listed in Tab.~\ref{tab1}. 
\begin{table}[b!]
\caption{The zero-temperature values of the sub-lattice magnetization and pair correlation functions inherent to the individual ground-state phases.}
\label{tab1}
\centering
\begin{tabular*}{\columnwidth}{@{\extracolsep{\fill}}p{5mm}ccccc}
\hline
 & $m_I$ & \hspace{-0.15cm}$m_{e}$ &  \hspace{-0.15cm}$C_{Ie}^{zz}$ &\hspace{-0.3cm}$C_{ee}^{zz}$ & \hspace{-0.3cm}$C_{ee}^{xx}$\\
\hline\hline
FM & $\frac{1}{2}$ & \hspace{-0.15cm}$1$ & \hspace{-0.15cm}$\frac{1}{6}$ & \hspace{-0.3cm}$\frac{1}{12}$ & \hspace{-0.3cm}$\frac{1}{12}$\\[0mm]
FRU& $0$ & \hspace{-0.15cm}$0$ & \hspace{-0.15cm}$0$ & \hspace{-0.3cm}$-\frac{1}{24}\left(1{+}\frac{U+2t}{\sqrt{U^2+4Ut+36t^2}}\right)$ &\hspace{-0.3cm}  $\frac{1}{12}\left(1{+}\frac{U+18t}{\sqrt{U^2+4Ut+36t^2}}\right)$ \\
\hline
\end{tabular*}
\end{table}
The saturated values of both the sub-lattice magnetization $m_I$, $m_e$ together with the positive values of the correlation functions $C_{Ie}^{zz}$, $C_{ee}^{zz}$ and $C_{ee}^{xx}$ in the first line of Tab.~\ref{tab1} clearly confirm a perfect spontaneous ferromagnetic spin order inside and between the Ising and electron sub-lattices, as well as the constant ferromagnetic correlation between mobile electrons from the same triangular cluster in the transverse direction across the whole FM phase. 
Different situation can be observed in the FRU phase. The zero magnetization $m_I$ points to the frustration of the Ising sub-lattice, while $m_e=0$ and the analytical expression of the correlation function $C_{ee}^{zz}$ are an evidence of the antiferromagnetic correlation between mobile electrons from the same triangular cluster along the $z$-axis. Contrary to this, $C_{ee}^{xx}$ indicates the existence of ferromagnetic transverse (quantum) correlation of these particles. It is obvious from density plots of both the correlation functions plotted in Fig.~\ref{fig:2} that the antiferromagnetic longitudinal correlation between electrons strengths upon increasing $U/J$ at the expense of the ferromagnetic transverse one in the FRU phase. Moreover, the zero-temperature asymptotic values (formulas) of the physical quantities in Tab.~\ref{tab1} clearly indicate that only the longitudinal pair correlation function $C_{ee}^{zz}$ corresponding to the FRU phase takes negative values for any combination of the parameters $t/J$ and $U/J$. The negative value of $C_{ee}^{zz}$ ensures that neither the products of correlation functions running along the electron triangle and along two vertices of the same electron triangle and the nearest nodal lattice site occupied by the Ising spin (see the $j$-th bipyramidal plaquette marked by the red ellipse in Fig.~\ref{fig:1}) cannot acquire positive values in the FRU phase. With this in mind one may conclude that not only the Ising sub-lattice, but also the electron one is spontaneously frustrated when the FRU phase is the ground state, namely in terms of the frustration concept developed by R.~J.~V.~dos~Santos and M.~L.~Lyra~\cite{San92}.
\begin{figure}[t!]
	\centering
	\vspace{0mm}
	\includegraphics[width=0.95\columnwidth]{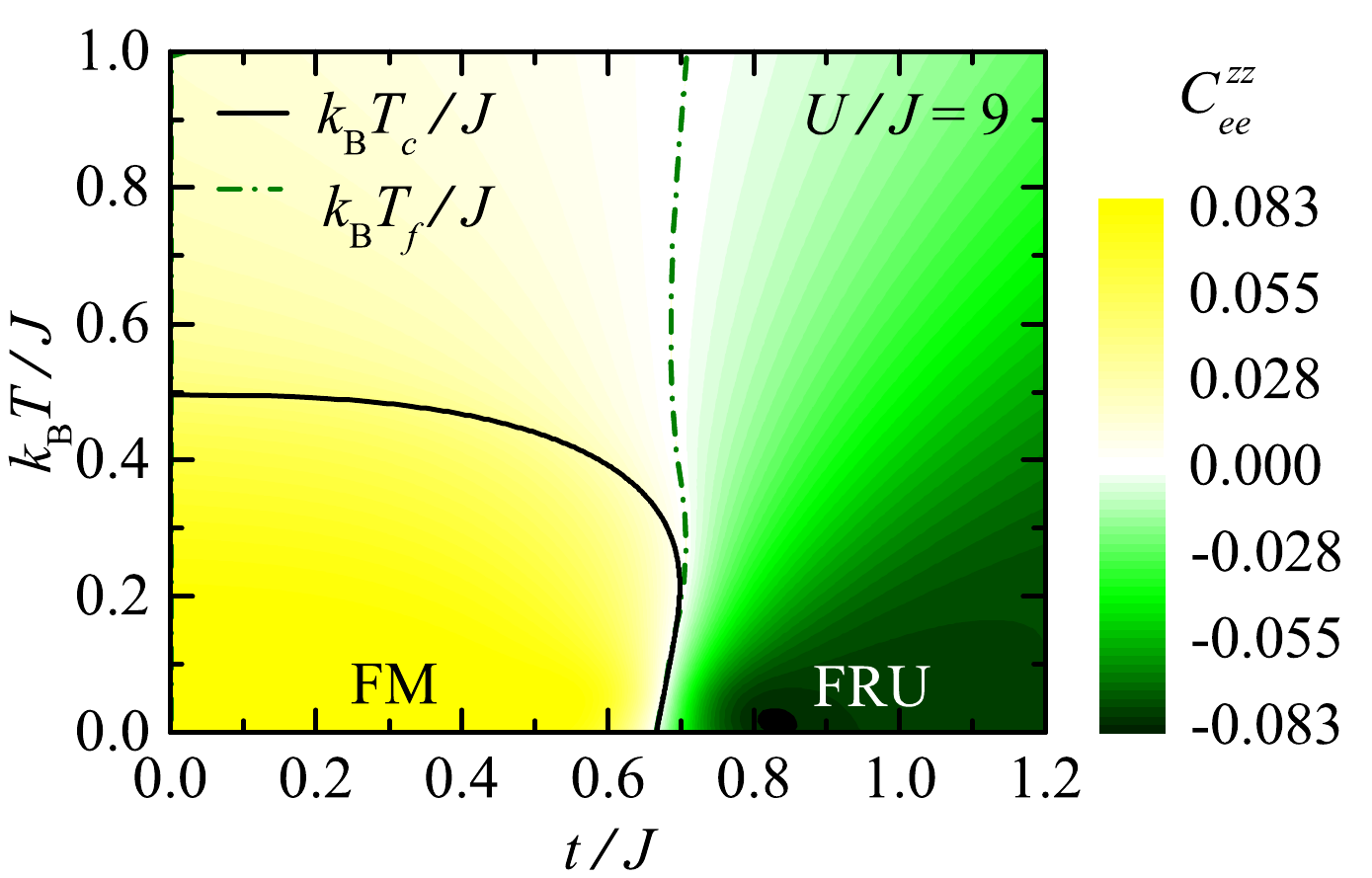}
	\vspace{-2mm}
	\caption{The density plot of the longitudinal pair correlation function specifying the spin ordering between two electrons in a given triangular cluster along $z$-axis in the $t/J-k_{\rm B}T/J$ plane for the Coulomb repulsion $U/J=9$. The green dash-dotted line indicates the frustration temperature and the black solid line shows the critical temperature of the model.}
	\label{fig:3}
\end{figure}

\subsection{Spin frustration at finite temperatures}
\label{subsec:2}

The non-chiral spin frustration of the electron sub-lattice identified in the disordered FRU phase persists also at finite temperatures. All important information on its thermal evolution can be read from Fig.~\ref{fig:3}, which shows the density plot of the correlation function $C_{ee}^{zz}$ in the $t/J-k_{\rm B}T/J$ plane for $U/J=9$. In this figure, the green dash-dotted and black solid lines indicate the frustration temperature $k_{\rm B}T_{\!f}/J$ delimiting the spin frustration in the electron sub-lattice and the critical temperature $k_{\rm B}T_{c}/J$ (continuous phase transition) of the model, respectively. The former temperature was numerically determined by setting $C_{ee}^{zz}=0$, while the latter one is the numerical solution of the exact critical condition $\sinh^2\!\big[J_{eff}/(k_{\rm B}T_c)\big] = 1$ for the spin-$1/2$ Ising square lattice~\cite{Ons44}. Evidently, the non-chiral spin frustration of the electron sub-lattice present in the FRU ground state does not interfere in any way with the region delimited by the critical temperature $k_{\rm B}T_c/J$, where the system is spontaneously ordered. In general, the phenomenon is suppressed by the temperature until it finally disappears at a certain $k_{\rm B}T_{\!f}/J$. The opposite trend can be observed only near the ground-state phase transition FRU--FM for $t/J\gtrsim \big[\!\sqrt{(U/J + 6)^2 + 24U/J} - U/J\big]/18$. In this particular region the frustration temperature $k_{\rm B}T_{\!f}/J$ shows a notable but not very pronounced S-shaped dependence merging with $k_{\rm B}T_c/J$ at low enough temperatures. Thus, one can find here up to three consecutive sign changes of the correlation function $C_{ee}^{zz}$, indicating a complete temperature-induced termination of the frustrated (anti-correlated) spin arrangement in the electron sub-lattice, its re-appearing and definitive suppression. 
\begin{figure}[th!]
	\centering
	\vspace{0mm}
	\includegraphics[width=0.95\columnwidth]{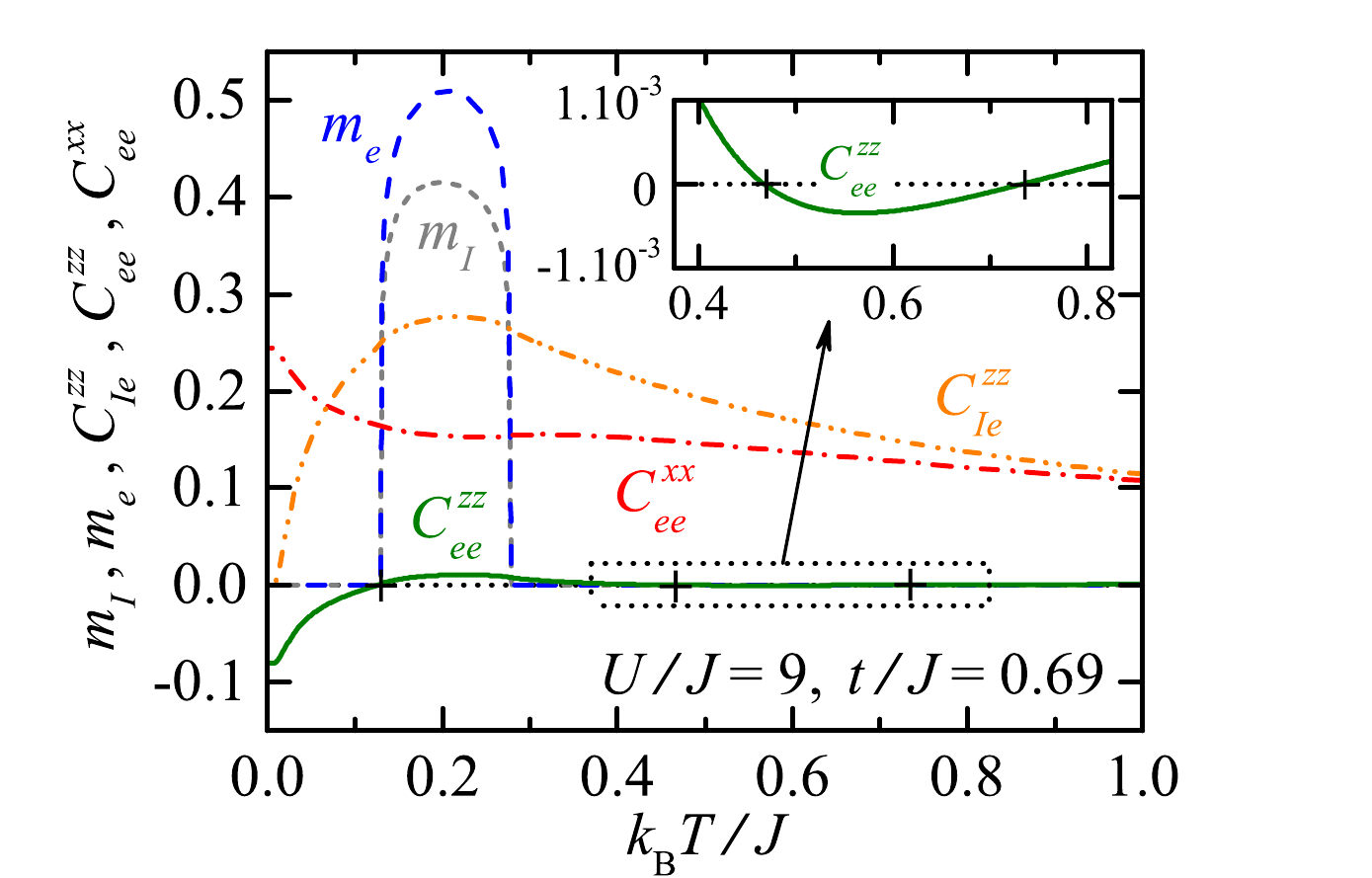}
	\vspace{-2mm}
	\caption{The temperature variations of the spontaneous sub-lattice magnetization and pair correlation functions corresponding to mobile electrons in the same triangular cluster of a given bipyramidal plaquette for the Coulomb repulsion $U/J=9$ and the hopping term $t/J=0.69$. The positions of three black crosses indicate three frustration temperatures of the system.}
	\label{fig:4}
\end{figure}
It can be deduced from Fig.~\ref{fig:3} that the first termination of the frustration is associated to the reentrant critical behaviour of the system due to thermal  activation of the long-range ferromagnetic spin-electron order peculiar to the neighbouring FM phase above the disordered FRU ground state. The above statements are in a good agreement with temperature dependencies of the sub-lattice magnetization $m_I$, $m_e$ and the pair correlation functions $C_{Ie}^{zz}$, $C_{ee}^{zz}$, $C_{ee}^{xx}$ depicted in Fig.~\ref{fig:4} for $U/J=9$ and $t/J=0.69$. 

\section{Conclusions}
\label{sec:4}

In the present paper, we have examined the frustration phenomenon in the exactly solvable spin-electron model on 2D lattice formed by identical bipyramidal plaquettes. It has been found that the ground state of the ferromagnetic version of the model contains the unfrustrated spontaneously ordered quantum ferromagnetic phase with two opposite chiral degrees of freedom of each electron triangular cluster and the disordered quantum one, where both the Ising and electron sub-lattices are frustrated. It has been demonstrated that the frustration of the electron sub-lattice persists even at finite temperatures, but only in the disordered region. In addition, the interesting thermally induced reentrant behaviour of the phenomenon with up to three consecutive frustration temperatures due to a thermally induced competition with the unfrustrated long-range ferromagnetic order has also been identified when the hopping term is selected from a close vicinity of the ground-state phase transition.


\end{document}